\documentclass[11pt,twoside]{article}

\usepackage{asp2006}
\usepackage{epsf}
\usepackage{psfig}
\usepackage{lscape}
\usepackage{graphicx}

\begin{document}
\title{Jet Properties and Evolution in Small and Intermediate Scale Objects}
\author{Marcello Giroletti} \affil{INAF Istituto di Radioastronomia, via
Gobetti 101, 40129, Bologna, Italy}

\begin{abstract}
Kinematic and spectral studies are improving our knowledge of the age
distribution in compact radio sources, providing evidence that small sources
are generally very young. The properties of jets in objects spanning the size
range from a few tens of parsecs to some kiloparsecs become then of particular
interest. Because of our selection criteria and of the small scales involved,
the properties of jets in the population of Compact Symmetric Objects (CSO) are
not well known yet. Polarization properties seem to indicate a strong influence
by the interaction with the dense surrounding medium, and some objects show
evidence of relativistic bulk motion. More evolved jets are present in the
class of Low Power Compact (LPC) sources and a number of cases are discussed
here. Since it is becoming increasingly clear that not all these sources will
survive to evolve into large scale radio galaxies, the question of the final
evolution of the CSO and LPC population is also discussed, with examples of
candidate dying sources.
\end{abstract}

\section{Introduction} 

\begin{figure}
\includegraphics[height=0.3\textwidth]{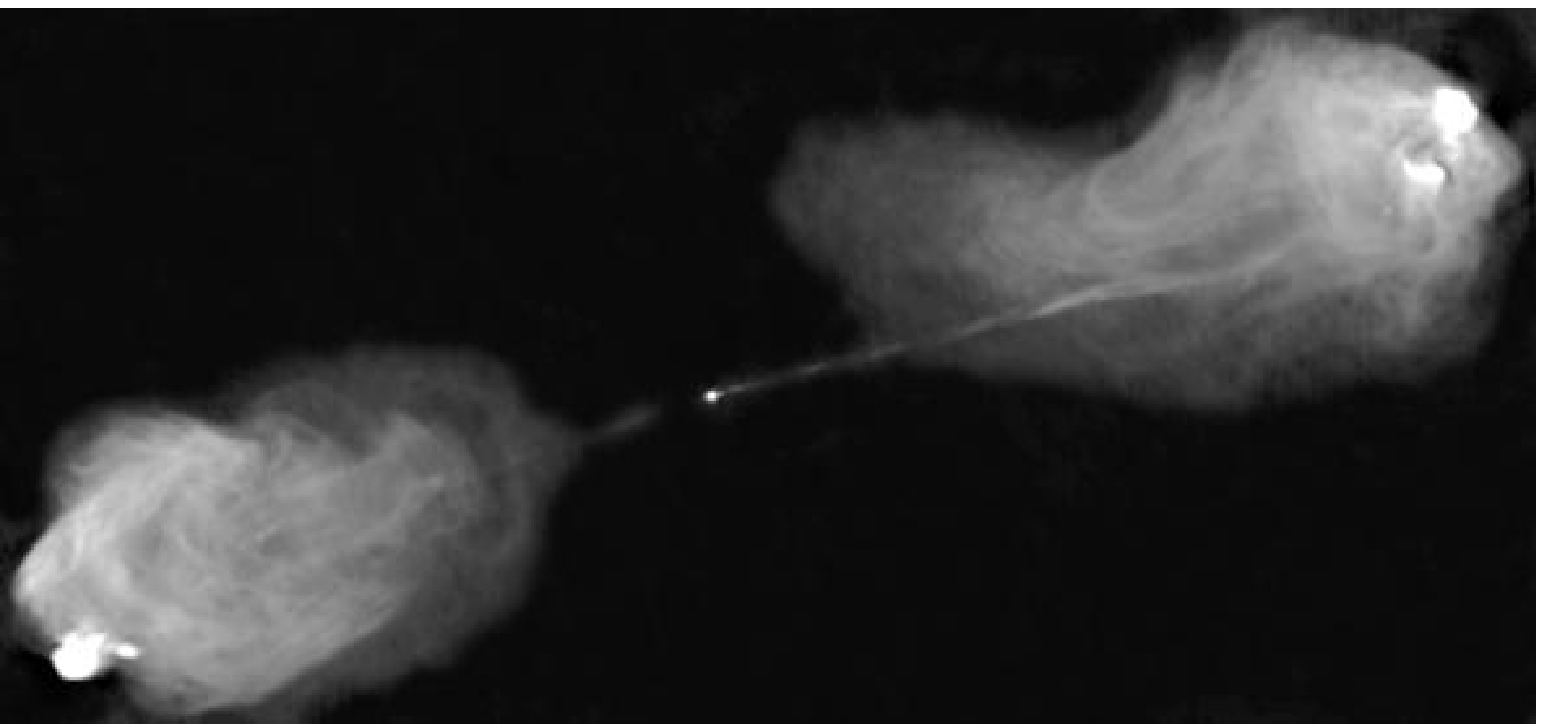}
\includegraphics[height=0.3\textwidth]{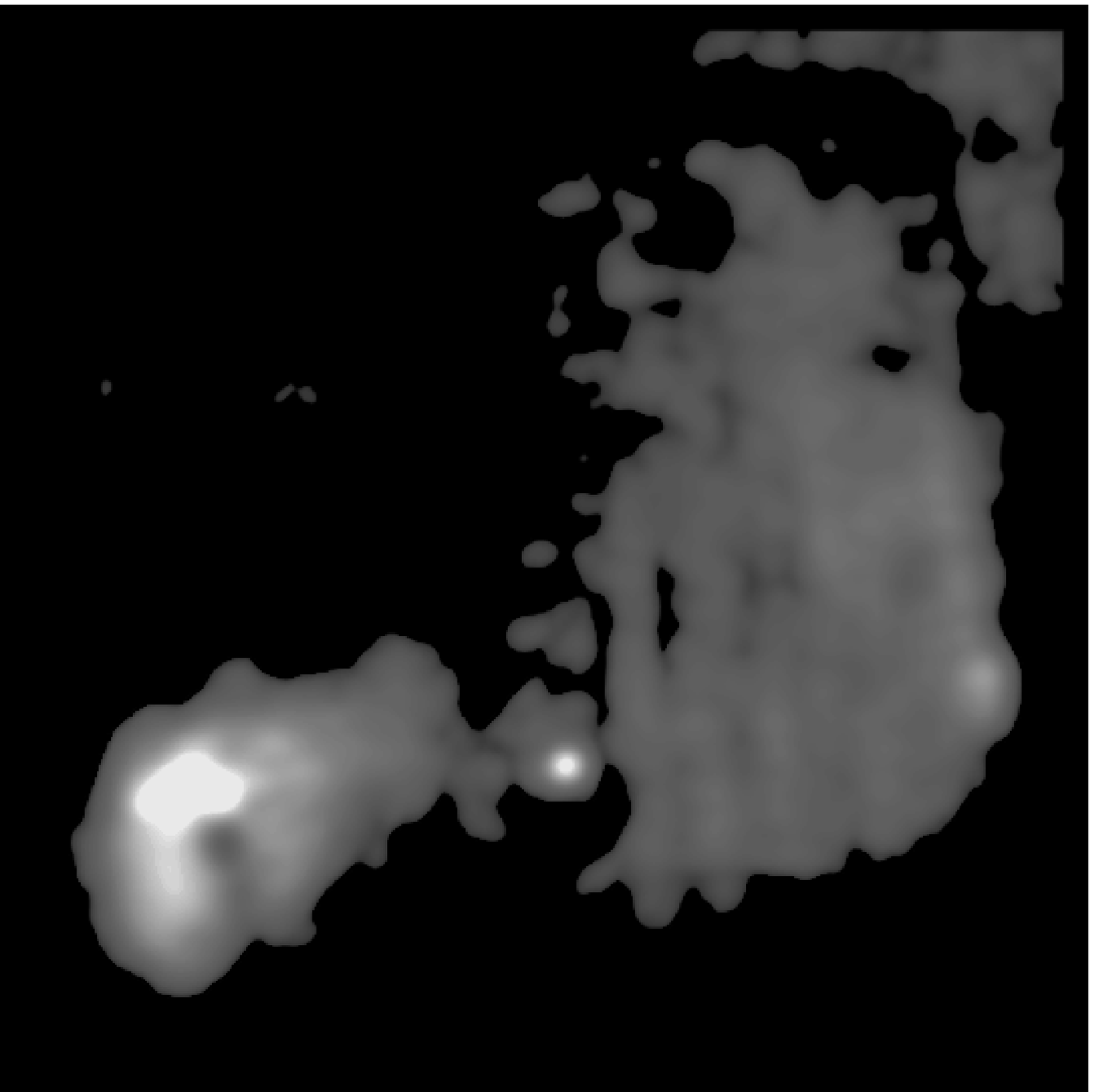}
\caption{ Left: Cygnus A, VLA 5 GHz (image courtesy of NRAO/AUI). Right:
4C31.04, Giroletti, VLBA 5 GHz. If brought to the same scale as Cygnus A,
4C31.04 would be only $\sim 1$ mm wide.} \label{f.1}
\end{figure}

Despite the large range of resolutions and scales sampled by different radio
telescopes and arrays, observations tend to ubiquitously reveal a common basic
structure. A central compact core is at the base of two jets flowing in
opposite directions, and forming diffuse lobes; bright hot spots can be present
at the site of interaction between the jet and the medium confining the
lobes. As Fig.~\ref{f.1} shows, this basic structure is found both on sources
that are extended over several hundreds kiloparsecs (such as Cygnus A, left
image) and in sources that are only a few tens parsecs (e.g.\ 4C31.04, right
picture).

An intriguing explanation of this observational fact consists of a self-similar
evolution of radio sources. Radio galaxies begin their life as (sub-)parsec
sized objects and then grow up to the kiloparsec scales, making their way
through the inter-stellar and inter-galactic medium (ISM and IGM,
respectively). In this process, the basic elements are maintained, and their
relative weights remain constant.

A number of refinements are required in order to improve over this basic
picture, and in particular to overcome the problem of the too large fraction of
small sources found in radio catalogs.

\begin{enumerate}

\item Not all small looking sources are physically small. Large radio sources,
when projected under a small viewing angle, will look rather compact. Because
of Doppler beaming, the fraction of such objects is even increased,
particularly in high frequency catalogs. Unified schemes \citep[see
e.g.][]{urr95} have been successful in posing a connection between compact and
extended sources on the basis of geometry; blazars are a clear example of this.

\item Not all physically small sources grow all the way to Mpc scale;
frustration from an over-dense external medium or a lack of fuel for the
central engine can be responsible of a premature end of the evolution from
small to large scale. While the evidence for frustration has not been found so
far, the search for examples of short-lived activity seems to find some
support, both on small and intermediate scales \citep[see
e.g.][]{gug05,kun05,kun06}

\item The total radio power $P$ of each object, after an initial increase with
radio size, falls as a power law function of linear size $LS$: $P \propto
(LS)^{-h}$. Differently from their compact counterparts, large radio galaxies
tend to escape flux limited searches. Models describing different evolutionary
tracks are proposed by \citet{beg96,ale00,sne00,tin06}.

\end{enumerate}

In the following sections, after some basics about radio source age estimates
(\S 2), we discuss the properties of jets in the case of the smallest radio
sources (\S 3) and of the more evolved low power compact sources (\S 4). We
touch upon the fate of these sources in \S 5 and give our final remarks in \S
6.

\section{Age estimates}

\subsection{Kinematics}

A classic way to estimate the age of a source lies in measuring the increase
$\Delta s$ in the separation between its outermost edges over some interval
$\Delta t$. From these values, one derives a mean advance velocity
$v_\mathrm{sep}= \Delta s / \Delta t$ and, assuming that $v_\mathrm{sep}$ has
been constant over the source lifetime, the ``0-order'' estimate of its age as
$t_\mathrm{kin}=LS/v_\mathrm{sep}$.

For a reference, let $v_\mathrm{sep}= 0.3c$ and $z=0.05$; the corresponding
$\Delta s / \Delta t$ is 0.1 mas yr$^{-1}$. Therefore, VLBI observations are
the only mean to obtain an estimate of $t_\mathrm{kin}$. Long time intervals,
repeated observations, and high frequency are important element to reveal the
motion and make a reliable age estimate.

The first successful report of an increase in hot spot separation was reported
in 0710+439 by \citet{ows98}. Several other studies have reported velocities of
$\sim 0.1-0.3 c$ in other sources, corresponding to ages in the range $\sim
10^2-10^4$ yr \citep{pol03,gir03,gug05}. 

\subsection{Spectral ages}

Alternatively, one can estimate the age of the source from the study of the
high frequency steepening in the spectrum of the emitting particles
population. Basic assumptions in the synchrotron theory are as follows:

\begin{itemize}
\item a power law initial energy distribution: $N(E)=N_0 E^{-\delta}$
\item radiative losses only: $dE/dt = -b H^2E^2$
\item a continuous injection of fresh particles: $Q(t)=N E^{-\delta}$
\end{itemize}

The observed spectrum is then

$$
S(\nu) = \left\{
\begin{array}{ll}
\nu^{-\alpha} & \mathrm{if\ } \nu < \nu_\mathrm{br} \\
\nu^{-(\alpha+0.5)} & \mathrm{if\ } \nu > \nu_\mathrm{br} 
\end{array}
\right.
$$

with $\alpha=(\delta+1)/2$ and $\nu_\mathrm{br} \simeq 10^9 \times
t_\mathrm{spec}^{­2} \times H^{­3}$ (GHz, yr, mG). Multifrequency observations
constrain $\nu_\mathrm{br}$ and allow us to derive $t_\mathrm{spec}$, if an
estimate of the magnetic field is available (equipartition conditions are
typically assumed).

Generally, $t_\mathrm{spec}$ estimates are in the range $10^3-10^4$ yr and they
agree quite well with kinematic ones \citep{mur99,ori07}. However, it is
important to be careful both in the assumptions going into the model
(equipartition magnetic field, no re-acceleration, synchrotron losses only) and
in the observations (integrated spectrum, matched $(u,v)$ coverage), as
discussed also by Nagai et al.\ at these conference.

\section{The smallest radio sources}

\begin{figure} 
\plottwo{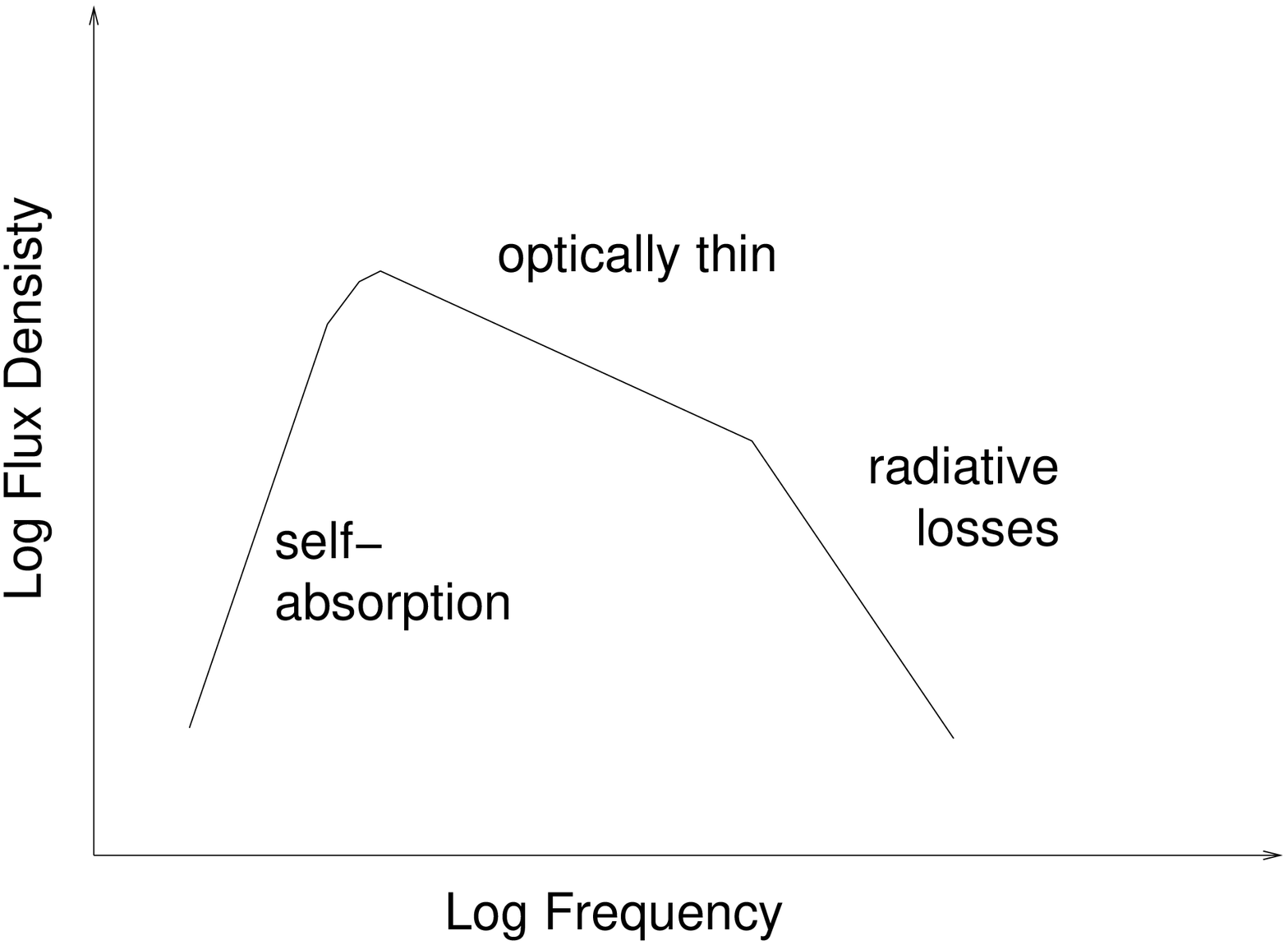}{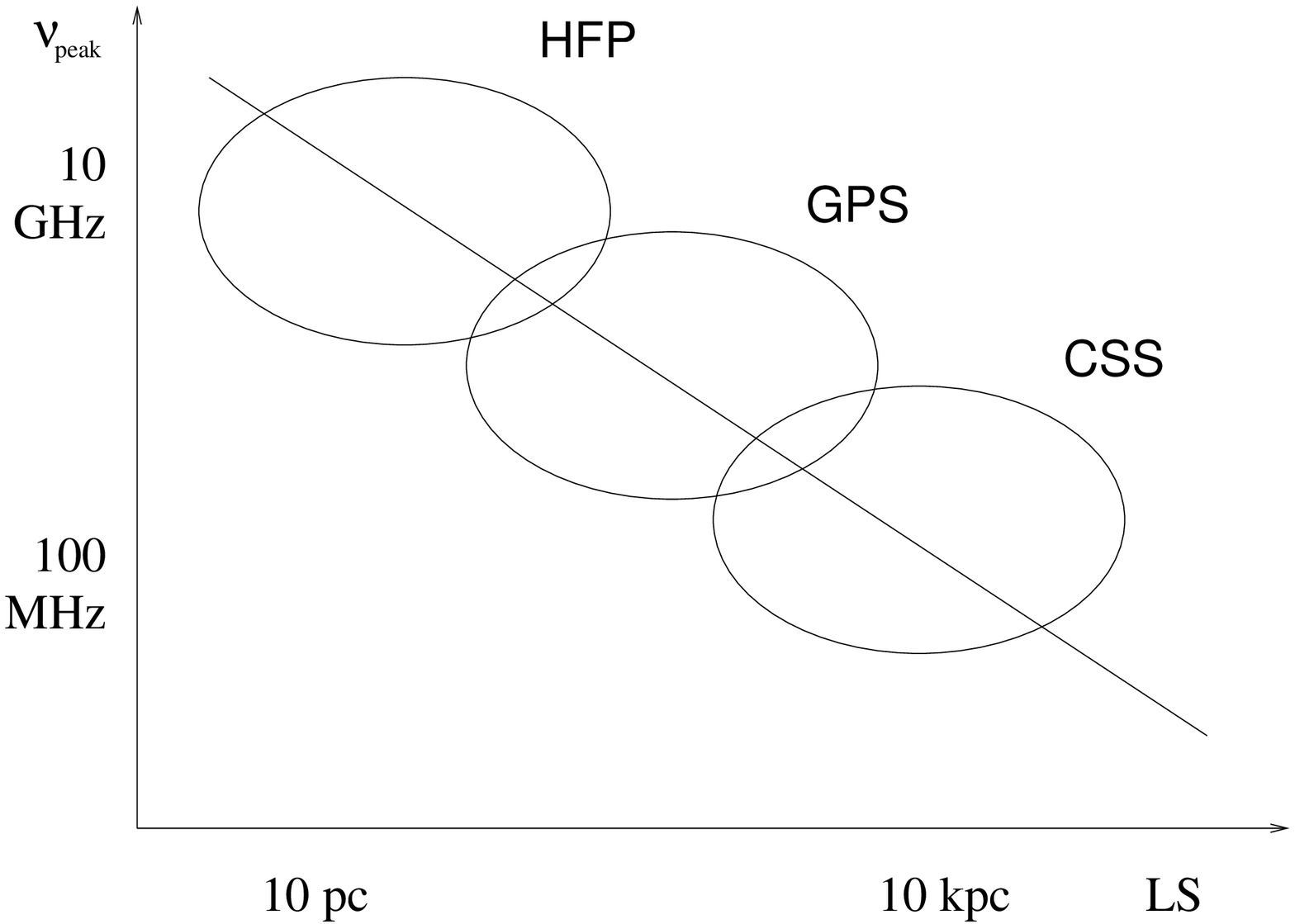}
\caption{Left: typical synchrotron spectrum; right: the HFP-GPS-CSS sequence in
the linear size vs turnover frequency plane.} \label{f.spectral}
\end{figure}

In the {\it youth scenario}, Compact Symmetric Objects (CSOs) represent the
earliest stage of the evolutionary path for radio galaxies. Their defining
properties can be outlined as follows:

\begin{enumerate}
\item {\bf Size/morphology.}  CSOs are typically sub-galactic in size ($LS <1 $
kpc); their structure, when resolved with VLBI, is that of either a symmetric
double or triple; flux density is usually quite stable
\item {\bf ISM related properties.} Probably because of the dense medium of the
region in which they form, CSOs tend to have low degree of polarization and to
show relatively high HI absorption
\item {\bf Radio spectrum.} Their spectrum is typical of synchrotron
self-absorbed radiation, i.e.\ convex with a power-law at high frequency and a
low frequency turnover. 
\end{enumerate}

These properties, and in particular the characteristic spectral shape, are
often used to improve the efficiency of the selection process (see
Fig.~\ref{f.spectral}).  Interestingly, an inverse correlation has been found
between linear size $LS$ and turnover frequency $\nu_\mathrm{peak}$
\citep{ode98}. The smaller the source, the higher the turnover frequency, so
that a spectral classification is also in use as follows: Compact Steep
Spectrum (CSS) sources, GHz Peaked Spectrum (GPS) sources, and High Frequency
Peakers \citep[HFP,][]{dal03}, with turnover frequency above 0.1, 1, and 10 GHz
respectively.  It is also worth reminding that blazar contamination becomes
more relevant as the $\nu_\mathrm{peak}$ increases. Variability and parsec
scale morphology are then required to distinguish between genuine HFPs and
blazars (see Orienti, these proceedings).

\subsection{Jets in CSOs}

\begin{figure}
\plottwo{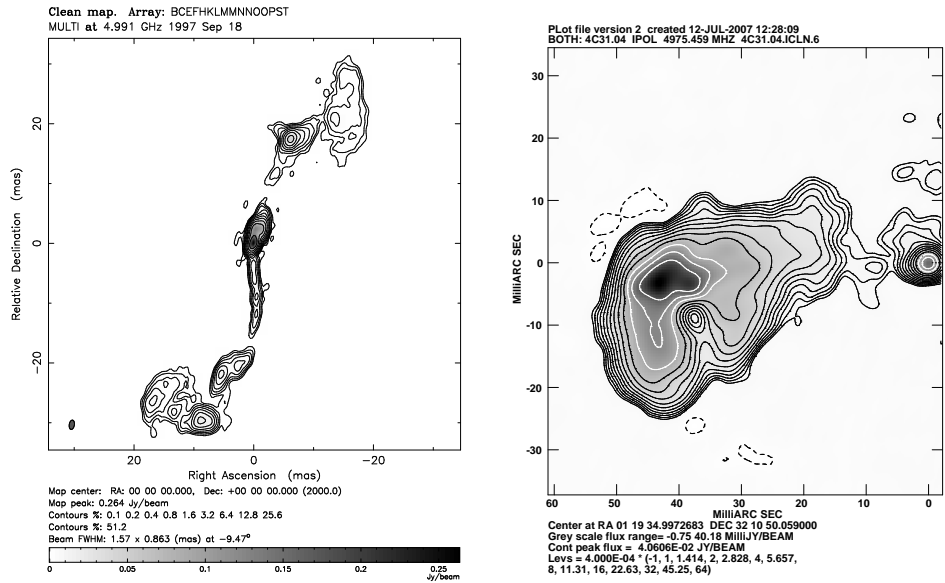}{giroletti3b.eps}
\caption{Left: a collimated jet is nicely visible in the 5 GHz image of the CSO
2353+495 \citep{ows99}. Right: the eastern lobe of 4C31.04, with the sharp bend
in the hot spot region, and the remarkable ``hole'' just next to it.}
\label{fig3}
\end{figure}

Since CSOs are selected on the basis of symmetry in arm length and flux
density, we are somehow biased in favor of sources in the plane of the
sky. This favors viewing angles $\theta \sim 90^\circ$, which causes the
Doppler factor $\delta = [\Gamma (1-\beta \cos \theta)]^{-1}$ of relativistic
jets to become $<1$. Therefore, relativistic jets in CSOs are typically
debeamed and thus hard to study. Moreover, the angular scales of CSO jets can
only be resolved with VLBI and, as a consequence, our knowledge of CSO jets as
a class is still relatively poor. There are however several objects in which
collimated jets are clearly detected \citep[e.g. 2353+495,
Fig. \ref{fig3},][]{ows99}. In general, the properties of the jets are strongly
influenced by the interaction with the peculiar medium surrounding these
sources.

\subsubsection*{Bending.}

Sharp bends in the jets are not uncommon and they are likely related to the
impact on denser clouds which deflects the jet. In the case of 4C31.04, a
``dentist drill'' model could also be at work \citep[see right panel of
Fig. \ref{fig3} and][]{gir03}

\subsubsection*{Proper motion}

Multi-epoch observations aimed at studying hot-spot advance can also be
fruitful in revealing component motions in the jet. While hot spot advance
velocities generally do not exceed $\sim 0.1c-0.3c$, jet components can have
somehow larger values.  Knots in 2353+495 have apparent velocities in the range
$0.3c-0.8c$ \citep{tay00}, which are quite typical for CSO jets. Subluminal
velocities are expected if the viewing angles are close to 90$^\circ$, as in
most CSOs. However, mildly superluminal velocities have been found, e.g. in
J1915+6548 \citep{gug07}. This source must therefore be more closely aligned to
our line of sight, as other properties also suggest (see below). In any case,
it seems that CSOs can have relativistic jets ($\gamma = 5-10$), once that the
advancing hot spots have cleared the densest ISM off the way.

\subsubsection*{Polarization}

\begin{figure}
\includegraphics[width=0.98\textwidth]{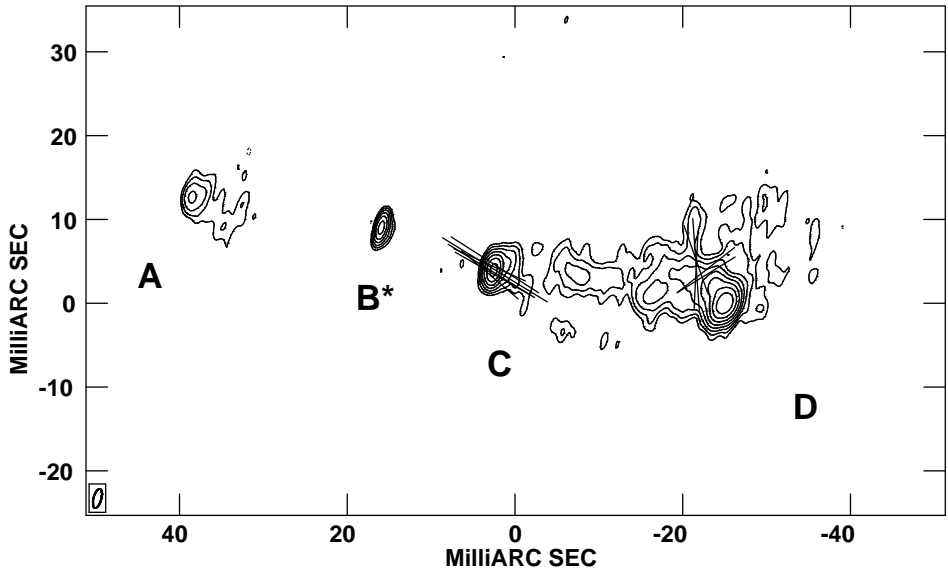}
\caption{Polarization in the jet of the CSO J1826+1831: fractional polarization
of component C is 9\%. B* is the core \citep[from][]{gug07}.}
\label{fig4}
\end{figure}

Synchrotron radiation is typically highly polarized. However, because of the
Faraday screen constituted by the rich medium around them, CSOs do not usually
show polarized flux. \citet{gug07} have recently found exceptions to this rule
in two CSOs that are less symmetric than the typical CSOs. In particular, the
strongest polarization in a CSO jet (9\%) has been found in J1826+1831, whose
arm length ratio is $\sim 3$ (see Fig.\ \ref{fig4}). Interestingly,
\citet{gug07} also report a Faraday rotation measure as low as $-180 \pm 10$
rad m$^{-2}$, which can be attributed to a shorter path length through the
circumnuclear torus and, consequently, a lower Faraday depth. Put altogether,
this evidence points to a jet closer to on-axis than in other CSOs, just as in
J1915+6548 whose jet components are both polarized and (weakly) superluminal.


\subsubsection{High energy emission} 

Besides the synchrotron radiation, there are other emitting processes that can
make CSO detectable in higher energy bands. In particular, Stawarz et al.\
(these conference) have suggested that CSOs could be strong $\gamma-$ray
sources thanks to the IC up-scattering of UV photons from the accretion disk
within the radio lobes. It will be intriguing to test this prediction with
GLAST.

\section{From CSOs to LPCs}

\begin{figure}
\plotone{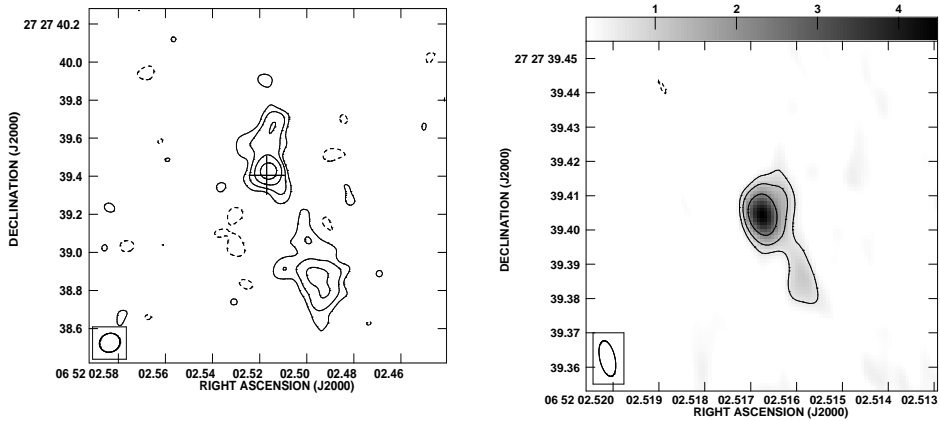}
\caption{Left: 22 GHz VLA image of 0648+37; the cross denotes the position of
the VLBA 1.6 GHz core/jet structure, shown in the right panel.}
\label{f.0648}
\end{figure}

On scales of a few kiloparsecs, we find sources that could be evolved CSOs. We
name them Low Power Compact sources, and we characterize them by their low
radio power ($P_\mathrm{LPC} \la 10^{24} - 10^{25}$ W Hz$^{-1}$) and linear
sizes around a few kpc. Thus, these sources are unresolved in low frequency
surveys such as the 3C and the B2, and even in most common VLA
configurations. Multiple causes can be invoked to explain the properties of
sources in this class: youth, instabilities in the jets, frustration, a
premature end of nuclear activity, or just a very low power core
\citep{gir05b}.

Only with observations at high frequency in the A configuration of the VLA, or
in phase referencing mode with VLBA, we can resolve LPC sources and discuss
their nature. We find rich substructures, including active jets, lobes, and
occasionally hot spots. These sources resemble, on an intermediate scale, both
the larger classical radio galaxies and the smaller CSOs; however, they tend to
be more frequently edge dimmed than higher power sources.

A remarkable LPC source is 0648+27, in which different instruments have been
exploited to reveal episodes of activity on different timescales
\citep{gir05b,emo06}. First off, the radio VLA data at 22 GHz resolve the
double source visible at lower frequency, revealing a compact component in the
northern lobe (Fig.~\ref{f.0648}, left panel). Phase referenced 1.6 GHz VLBI
data reveal that this component is actually a flat spectrum core
($\alpha^{22}_{1.6} = 0.47 \pm 0.03$), at the base of a faint jet
(Fig.~\ref{f.0648}, right). The source does not have advancing hot spots and it
is not possible to estimate a kinetic age as in typical, young, CSOs. Its
larger size fits the LPC definition best, and its spectral age estimate is
$\sim 1$ Myr.  Neutral hydrogen 21 cm observations bear also the signature of a
major merger occurred $\sim 1.5$ Gyr ago, while optical spectroscopy suggests
an event of starburst activity $\sim 0.3$ Gyr. Although it is difficult to
understand possible connections between the onset of the radio activity and the
other events, it is important to gather information across the whole spectrum.

\begin{figure}
\begin{center}
\plottwo{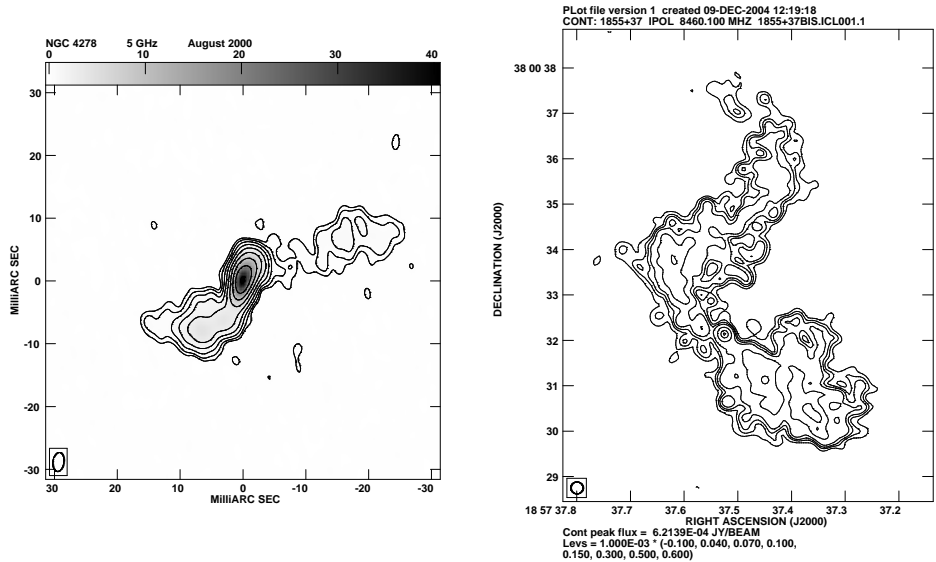}{giroletti6b.eps}
\caption{Left: a 5 GHz VLBA image of the two sided parsec scale structure of
the LLAGN NGC4278 \citep{gir05a}. Right: 8.4 GHz image of the head-tail LPC
1855+37, a good candidate of a dying radio source \citep{gir05b}.}
\label{f.lpc}
\end{center}
\end{figure}

Another interesting source fitting the the LPC definition is NGC\,4278 (see
Fig.~\ref{f.lpc}, left panel); it has a 5 GHz luminosity of $\sim 10^{22}$ W
Hz$^{-1}$, typical for LLAGNs, and it looks compact on kpc scale. VLBA
observations at 5 and 8.4 GHz resolve the source in a flat spectrum core
surrounded by two-sided pc scale jets. From a two epochs study, \citet{gir05a}
have revealed a proper motion with sub-relativistic velocity (highest apparent
speed $v=0.1c$), and estimated a Lorentz factor $1.2 < \Gamma < 1.7$ and a
viewing angle of a few degrees.

Several other LPCs have been studied \citep[and in prep.]{gir05b}, revealing
cases of jet instabilities, both in space (e.g.\ in 0222+36) and time (as in
0258+35). In general, high resolution, multi frequency radio observations seem
to be successful in understanding the properties of the jets of LPC sources,
which are typically $<1$ kpc long, often two-sided, and feed ``mini''-lobes,
rarely embedding hot spots. These jets seem to be in mildly relativistic
regime; on the other hand, the mean advance velocity $v_\mathrm{mean}$ of the
radio source, obtained by the simple ratio between the linear size and the
spectral age ($v_\mathrm{mean}=LS/t_\mathrm{spec}$), is typically
$v_\mathrm{mean}\ll c$.

\section{The fate of LPC sources}

The low value of the inferred advance velocities raises the question of the
possible future evolution of LPC sources. While only a fraction of them seems
capable of advancing further, and eventually form a large scale FRI or FRII
radio galaxy, many of these sources seem unable to form large kpc scale lobes
and they will thus face a different destiny. Intermittent activity of the
central engine is a possibility, as suggested by the large scale emission found
around some pc scale CSO (see e.g.\ Jamrozy et al., these conference). A
premature end of activity can take place in other sources, which are however
difficult to find, since $\nu_\mathrm{br}$ goes down very rapidly after the
injection of fresh particles stops. Only in denser media, e.g. in an X-ray
emitting cluster, the fossil phase could last long enough to facilitate our
search \citep{mur05}.

An LPC that could be a good example of dying radio source is 1855+37
\citep{gir05b}. The source has significant extended emission at low frequency
($S_{408} = 600$ mJy) and is resolved in a nice head-tail at 8.4 GHz
(Fig.~\ref{f.lpc}, right panel). However, the VLA core is much weaker than
expected ($S_C = 0.6$ mJy), even if we account for a strong Doppler de-beaming,
and it is not detected with VLBA at 1.6 GHz, suggesting that the nuclear
activity might be going off. Moreover, the whole source is also not detected at
$\nu > 8.4$ GHz, which could be due to the lack of fresh electrons in the
lobes. Finally, the source is in cluster, which could have prevented it from
fading too rapidly and escaping detection even at low frequency.

Low frequency observations are particularly important to discover dying
sources.  \citet{par07} have extracted LPCs with very steep spectra
($\alpha_{0.33}^{1.4} > 1.3$) from a cross correlation of the WENSS and the
NVSS surveys. Follow-up VLA multi-frequency observations have confirmed that a
number of them could be dying and/or restarted sources.

\section{Conclusions} 

The study of extragalactic jets is still posing a number of unanswered
questions (e.g.\ Blandford, this conference). Small and intermediate scale
objects can help to answer some important ones. In particular, they can be
relevant in the following issues:

\begin{itemize}

\item CSOs are the ideal targets to look for jets first steps, and they can be
exploited to investigate the reasons of the onset of the radio activity.

\item CSO jets strongly interact with the surrounding dense medium and can help
to understand its properties, as shown by their significant bends and strong
de-polarization. However, because of our current selection criteria, which
gives rise to Doppler de-beaming, and of their intrinsic compactness, jets in
CSOs are still difficult to study.

\item LPCs have more evolved jets, probably relativistic, but not always
capable of forming 100's kiloparsecs lobes. The reasons of this fact are not
well understood yet but they could be related to the process of accretion on
the central super massive black hole.

\item A premature end of the nuclear activity is among the possible
explanations invoked for the above fact. Though usually difficult to discover,
dying or restarted compact radio sources are interesting targets of recent
observations.

\end{itemize}

\acknowledgements We thank the organizing committees for making this excellent
conference possible. The National Radio Astronomy Observatory is a facility of
the National Science Foundation operated under cooperative agreement by
Associated Universities, Inc. This material is based in part upon work
supported by the Italian Ministry for University and Research (MIUR) under
grant COFIN 2003-02-7534.


\begin{thebibliography}{}
\bibitem[Alexander(2000)]{ale00} Alexander, P.\ 2000, \mnras, 319, 8
\bibitem[Begelman(1996)]{beg96} Begelman, M.~C.\ 1996, Cygnus A -- Study of a
Radio Galaxy, 209
\bibitem[Dallacasa(2003)]{dal03} Dallacasa, D.\ 2003, Publications of the
Astronomical Society of Australia, 20, 79
\bibitem[Emonts et al.(2006)]{emo06} Emonts, B.~H.~C., Morganti, R., Tadhunter,
C.~N., Holt, J., Oosterloo, T.~A., van der Hulst, J.~M., \& Wills, K.~A.\ 2006,
\aap, 454, 125
\bibitem[Giroletti et al.(2003)]{gir03} Giroletti, M., Giovannini, G., Taylor,
G.~B., Conway, J.~E., Lara, L., \& Venturi, T.\ 2003, \aap, 399, 889
\bibitem[Giroletti et al.(2005a)]{gir05a} Giroletti, M., Taylor, G.~B., \&
Giovannini, G.\ 2005, \apj, 622, 178
\bibitem[Giroletti et al.(2005b)]{gir05b} Giroletti, M., Giovannini, G., \&
Taylor, G.~B.\ 2005, \aap, 441, 89
\bibitem[Gugliucci et al.(2005)]{gug05} Gugliucci, N.~E., Taylor, G.~B., Peck,
A.~B., \& Giroletti, M.\ 2005, \apj, 622, 136
\bibitem[Gugliucci et al.(2007)]{gug07} Gugliucci, N.~E., Taylor, G.~B., Peck,
A.~B., \& Giroletti, M.\ 2007, \apj, 661, 78
\bibitem[Kunert-Bajraszewska et al.(2005)]{kun05} Kunert-Bajraszewska, M.,
Marecki, A., Thomasson, P., \& Spencer, R.~E.\ 2005, \aap, 440, 93
\bibitem[Kunert-Bajraszewska et al.(2006)]{kun06} Kunert-Bajraszewska, M.,
Marecki, A., \& Thomasson, P.\ 2006, \aap, 450, 945
\bibitem[Murgia et al.(1999)]{mur99} Murgia, M., Fanti, C., Fanti, R.,
Gregorini, L., Klein, U., Mack, K.-H., \& Vigotti, M.\ 1999, \aap, 345, 769
\bibitem[Murgia et al.(2005)]{mur05} Murgia, M., Parma, P., de Ruiter, H.~R.,
Mack, K.-H., \& Fanti, R.\ 2005, X-Ray and Radio Connections
(eds.~L.O.~Sjouwerman and K.K Dyer) Published electronically by NRAO,
http://www.aoc.nrao.edu/events/xraydio
\bibitem[O'Dea(1998)]{ode98} O'Dea, C.~P.\ 1998, \pasp, 110, 493
\bibitem[Orienti et al.(2007)]{ori07} Orienti, M., Dallacasa, D., \&
Stanghellini, C.\ 2007, \aap, 461, 923
\bibitem[Owsianik \& Conway(1998)]{ows98} Owsianik, I., \& Conway, J.~E.\ 1998,
\aap, 337, 69
\bibitem[Owsianik et al.(1999)]{ows99} Owsianik, I., Conway, J.~E., \&
Polatidis, A.~G.\ 1999, New Astronomy Review, 43, 669
\bibitem[Parma et al.(2007)]{par07} Parma, P., Murgia, M., de Ruiter, H.~R.,
Fanti, R., Mack, K.~-H., \& Govoni, F.\ 2007, ArXiv e-prints, 705,
arXiv:0705.3209
\bibitem[Polatidis \& Conway(2003)]{pol03} Polatidis, A.~G., \& Conway, J.~E.\
2003, PASA, 20, 69
\bibitem[Snellen et al.(2000)]{sne00} Snellen, I.~A.~G., Schilizzi, R.~T.,
Miley, G.~K., de Bruyn, A.~G., Bremer, M.~N., R\"ottgering, H.~J.~A.\
2000, \mnras, 319, 445
\bibitem[Taylor et al.(2000)]{tay00} Taylor, G.~B., Marr, J.~M., Pearson,
T.~J., \& Readhead, A.~C.~S.\ 2000, \apj, 541, 112
\bibitem[Tinti \& de Zotti(2006)]{tin06} Tinti, S., \& de Zotti, G.\ 2006,
\aap, 445, 889
\bibitem[Urry \& Padovani(1995)]{urr95} Urry, M.~E., \& Padovani, P.\ 1995,
\pasp, 107, 803
\end{thebibliography}
\end{document}